\documentclass[letterpaper]{article} 
\usepackage{aaai2026}  
\usepackage{times}  
\usepackage{helvet}  
\usepackage{courier}  
\usepackage[hyphens]{url}  
\usepackage{graphicx} 
\usepackage{natbib}  
\usepackage{caption} 
\usepackage{algorithm}
\usepackage{algorithmic}
\usepackage{booktabs}
\usepackage{tabularx}
\usepackage{url} 

\usepackage{xcolor}
\newcommand{\answerYes}[1]{\textcolor{blue}{#1}} 
\newcommand{\answerNo}[1]{\textcolor{teal}{#1}} 
\newcommand{\answerNA}[1]{\textcolor{gray}{#1}}

\usepackage{amssymb}

\usepackage{xcolor}
\usepackage{colortbl}
\definecolor{rowpurple}{HTML}{E6DDF0}
\definecolor{rowblue}{HTML}{E7F0FA}
\definecolor{roworange}{HTML}{F7E6CC}
\definecolor{rowyellow}{HTML}{FBF3D6}
\definecolor{sentimentRed}{RGB}{220,80,80}
\definecolor{sentimentGray}{RGB}{150,150,150}
\definecolor{sentimentGreen}{RGB}{80,200,120}

\newcommand{\Yes}{\textcolor{green!60!black}{$\checkmark$}}
\newcommand{\No}{\textcolor{red!70!black}{$\times$}}
\newcommand{\Partial}{\textcolor{yellow!70!black}{$\triangle$}}

\usepackage{newfloat}
\usepackage{listings}
\usepackage{subcaption}
\DeclareCaptionStyle{ruled}{labelfont=normalfont,labelsep=colon,strut=off} 
\floatstyle{ruled}
\newfloat{listing}{tb}{lst}{}
\floatname{listing}{Listing}
\title{SocialPulse: An Open-Source Subreddit Sensemaking Toolkit}
\author {
    Stephanie Birkelbach\textsuperscript{\rm 1},
    Maria Teleki\textsuperscript{\rm 1},
    Peter Carragher\textsuperscript{\rm 2},
    Xiangjue Dong\textsuperscript{\rm 1}*,\\
    Nehul Bhatnagar\textsuperscript{\rm 3}*,
    James Caverlee\textsuperscript{\rm 1}
}
\affiliations {
    \textsuperscript{\rm 1}Texas A\&M University,
    \textsuperscript{\rm 2}Carnegie Mellon University,
    \textsuperscript{\rm 3}Revionics\\
    \{birkelbachs, mariateleki, xj.dong, caverlee\}@tamu.edu,
    petercarragher@cmu.edu,
    nbhatnagar3010@gmail.com
}

\usepackage{bibentry}

\begin{document}

\maketitle

\begin{abstract}
Understanding how online communities discuss and make sense of complex social issues is a central challenge in social media research, yet existing tools for large-scale discourse analysis are often closed-source, difficult to adapt, or limited to single analytical views. We present SocialPulse, an open-source subreddit sensemaking toolkit that unifies multiple complementary analyses -- topic modeling, sentiment analysis, user activity characterization, and bot detection -- within a single interactive system. SocialPulse enables users to fluidly move between aggregate trends and fine-grained content, compare highly active and long-tail contributors, and examine temporal shifts in discourse across subreddits. The demo showcases end-to-end exploratory workflows that allow researchers and practitioners to rapidly surface themes, participation patterns, and emerging dynamics in large Reddit datasets. By offering an extensible and openly available platform, SocialPulse provides a practical and reusable foundation for transparent, reproducible sensemaking of online community discourse.
\end{abstract}

\begin{links}
    \link{Code}{https://github.com/birkelbachs/SocialPulse}
    \link{Video Demo}{https://youtu.be/zRprjXwFwQ0}
\end{links}

\section{Introduction}

Understanding how online communities discuss and make sense of complex social issues is a central challenge in social media research. Despite the scale of user-generated content online, prior work shows that analyses of this content often over-represent highly active users, obscuring the perspectives of long-tail contributors \cite{Oswald2025TipOfTheIceberg}. Tools that support analysis across participation levels are therefore critical for accurately characterizing online discourse.
At the same time, there is growing demand for fine-grained analysis of web discourse across domains, including public opinion monitoring, policy-making, and crisis response \cite{zhu2024study, yin-etal-2025-disastir}. Researchers and practitioners increasingly seek to identify latent themes, shifts in discussion, and patterns of participation in online conversations \cite{islam2025discovering}. Exploratory data analysis (EDA) tools that support rapid \textit{sensemaking} are particularly valuable in these settings:

\begin{quote}
    \textit{``Sensemaking is a motivated, continuous effort to understand connections (which can be among people, places, and events) in order to anticipate their trajectories and act effectively.''} - \citet{klein2006making} 
\end{quote}

\begin{figure}
    \centering
    \includegraphics[width=1\linewidth]{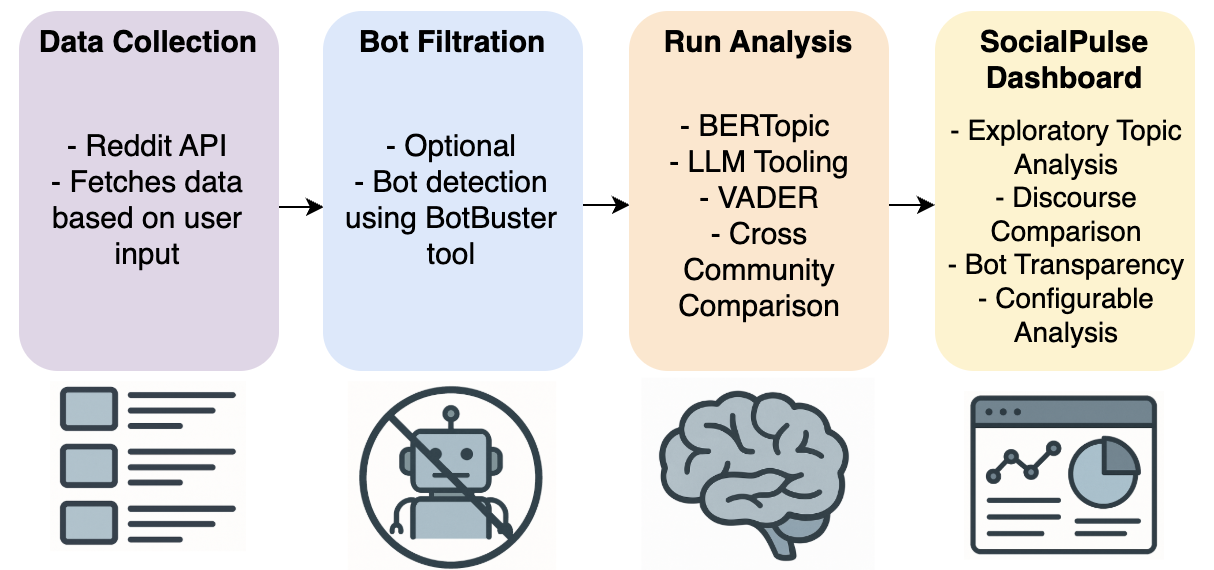}
    \caption{\textbf{The SocialPulse pipeline supports rapid exploratory data analysis and sensemaking of Reddit communities}; implementation details are provided in Table~\ref{tab:links}.}
    \label{fig:pipeline}
\end{figure}

\begin{figure*}[t]
    \centering
    \includegraphics[width=1\linewidth]{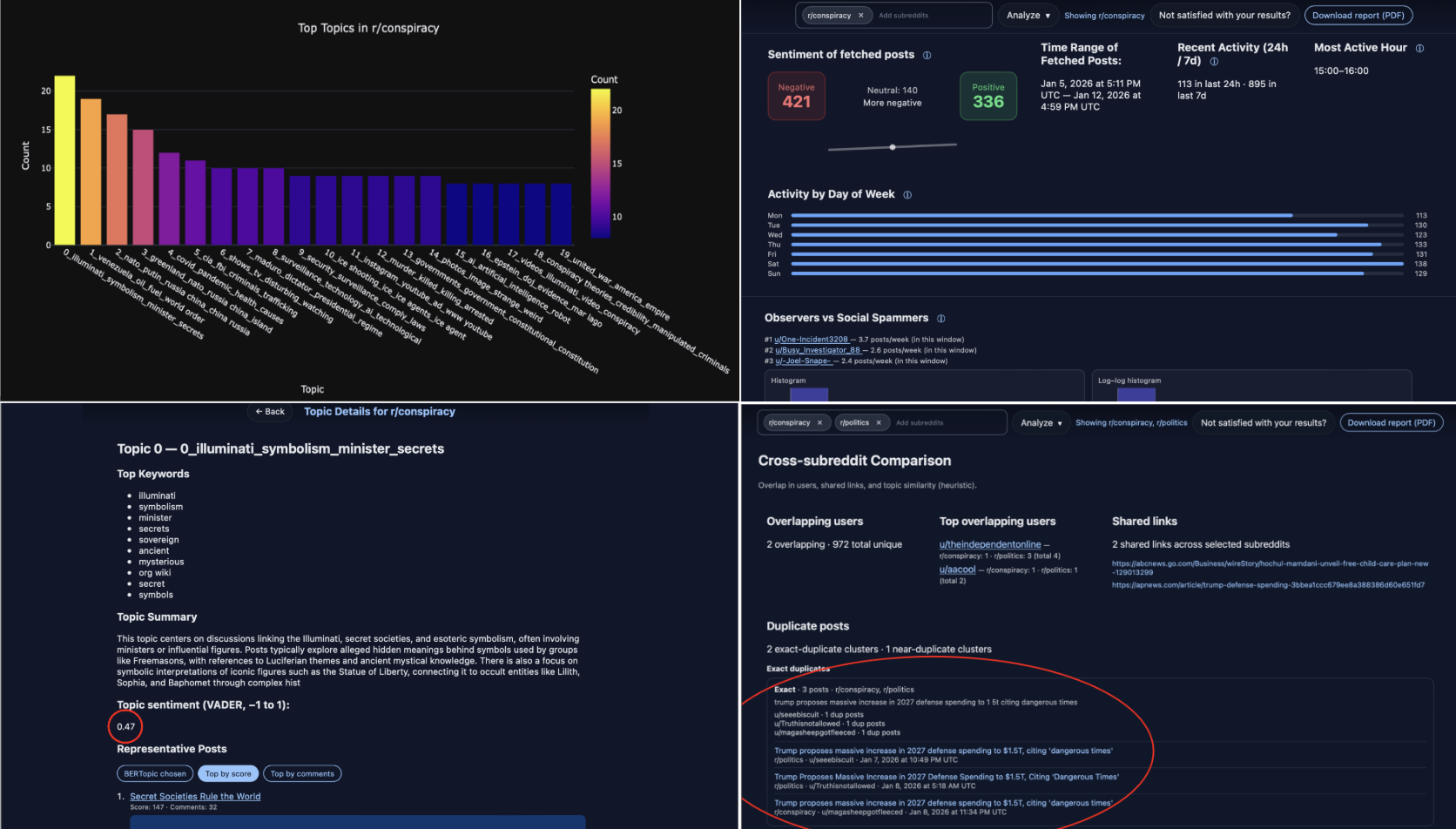}
    \caption{\textbf{The SocialPulse analytics interface supports rapid, multi-level sensemaking of Reddit discourse.} The interface enables (\textbf{a}, \textit{top left}) interactive topic exploration, (\textbf{b}, \textit{top right}) temporal and sentiment analysis, (\textbf{c}, \textit{bottom left}) topic-specific analysis, and (\textbf{d}, \textit{bottom right}) cross-subreddit comparison, illustrated here for \texttt{r/conspiracy} and \texttt{r/politics}.}
    \label{fig:dashboard}
\end{figure*}

Reddit is a widely studied platform for social science research, offering large-scale, community-structured discussions across diverse topics \cite{de2014mental, kumar2018community,dong-etal-2020-transformer,guo-etal-2020-benchmarking,davidson2023use}. Topic modeling has been central to analyzing such discourse, enabling the identification of latent themes and their temporal dynamics, with applications to gender norms \cite{teleki2025masculine}, sociopolitical conflict \cite{steffen2025more}, and more.
These works highlight that in online discourse, meaning emerges not only from topical content but also from participation structures, community norms, and interaction patterns. 
However, practical analysis remains challenging due to social bots, spam, and fragmented analytical systems \cite{ng2023botbuster, mendoza2024detection}, motivating the need for open-source, adaptable sensemaking tools that integrate topical, user-, and community-level perspectives (Table~\ref{tab:related-tools}).

To address these challenges, we present \textbf{SocialPulse}, an open-source subreddit sensemaking toolkit for interactive, exploratory analysis of Reddit discussions. 
We provide (i) an \textbf{overview of the analysis pipeline} (Figure~\ref{fig:pipeline}), (ii) \textbf{implementation details} (Table~\ref{tab:links}), and (iii) a \textbf{case study} yielding exploratory insights here. 


\section{SocialPulse System Overview}

SocialPulse supports exploratory sensemaking of Reddit communities through an integrated analysis pipeline (Figure~\ref{fig:pipeline}; Appendix~A) and an interactive dashboard for exploration and comparison (Figure~\ref{fig:dashboard}).
The architecture comprises four sequential stages—data ingestion, optional bot filtration, analytical modeling, and interactive visualization -- which together enable researchers to move fluidly between aggregate patterns and fine-grained content while accounting for participation dynamics and data quality concerns.

The system ingests Reddit data based on user-specified subreddits and configurable collection strategies, supporting both high-traffic and long-tail communities. An optional bot filtration stage allows users to identify and filter automated accounts, enabling analysis of human-driven discourse when desired. SocialPulse then applies a suite of complementary analytical methods to extract thematic, temporal, and behavioral patterns in Reddit discussions. Topic modeling is performed using BERTopic, which leverages transformer-based embeddings and clustering to identify latent themes in text, while VADER sentiment analysis is used to characterize the polarity of posts. The system further allows cross-community comparison, enabling the analysis of shared themes, overlapping users, and duplicated content. We provide the full implementation details in Appendix~A. 

Analysis results are presented via an interactive dashboard, which enables users to compare discourse across communities, examine topic structure and sentiment over time, and move from aggregate summaries to individual posts and comments. 

\section{Case Study: Sensemaking in \texttt{r/conspiracy}}


Online discussions of conspiracy theories present a challenging setting for sensemaking, as they often involve a heterogeneous mix of genuine belief, skepticism, and play \cite{samory2018conspiracies}. In this section, we demonstrate how SocialPulse supports sensemaking within and across \texttt{r/conspiracy}, yielding three exploratory insights.

Applying SocialPulse to \texttt{r/conspiracy} over the week of 1/4/26, BERTopic identifies 20 distinct topics with a high degree of thematic diversity as seen in \textbf{Figure~\ref{fig:dashboard}a}: the topic labels range from long-standing conspiracy topics (e.g., aliens, intelligence agencies, war, and consciousness) to discussions about current events. Topic frequencies are relatively evenly distributed, and most topics are well-separated with little overlap, suggesting clear thematic boundaries. \textbf{Figure~\ref{fig:dashboard}b} shows posting activity is at a high on Saturdays and at a low on Mondays, with the highest volume occurring from 15:00-16:00 UTC. Sentiment polarity across collected posts skews slightly negative (\textcolor{sentimentRed}{421 negative posts}, \textcolor{sentimentGray}{140 neutral posts}, \textcolor{sentimentGreen}{336 positive posts}). Finally, user activity over the week is dominated by long-tail participation where most users post less than once a week. 
As shown in \textbf{Figure~\ref{fig:dashboard}c}, specifically examining Topic 0 reveals that topic-level sentiment polarity is largely \textcolor{sentimentGray}{neutral} (\textcolor{sentimentGray}{score = 0.47}) within this topic. 
Looking individually at each topic (\textbf{unpictured}), we see that topics centered on current events exhibit a more \textcolor{sentimentRed}{negative} sentiment, while topics associated with long-standing conspiracy theories such as consciousness, religion and cults, and historical theories (e.g. the Illuminati or the moon landing) tend to be more \textcolor{sentimentGray}{neutral} or \textcolor{sentimentGreen}{positive}. Adjusting the data collection window to include posts from earlier weeks further reveals differences in topic sentiment over time: current event topics become more generalized and overlapping, while long-standing conspiracy topics remain relatively semantically stable. Extending this analysis to a cross-subreddit comparison with \texttt{r/politics} in \textbf{Figure~\ref{fig:dashboard}d}
reveals measurable overlap in both users and content, including two shared users, two identical external links posted in both subreddits, and multiple duplicated posts; all of the duplicated posts discussed current events. Furthermore, there are more social spammers in \texttt{r/conspiracy} than in \texttt{r/politics}, and the sentiment scores for both subreddits are predominantly \textcolor{sentimentRed}{negative}. 

\textbf{\textit{$\rhd$ Exploratory Insight 1: Topic-Dependent Sentiment Patterns}} Topic-level analysis reveals systematic differences in sentiment between discussion types within \texttt{r/conspiracy}. Topics centered on current events tend to exhibit more \textcolor{sentimentRed}{negative} sentiment and greater instability over time, while long-standing conspiracies remain relatively stable and are often more \textcolor{sentimentGray}{neutral} or \textcolor{sentimentGreen}{positive} in tone. 

\textbf{\textit{$\rhd$ Exploratory Insight 2: Temporal Variation in Community Engagement}} Posting activity within \texttt{r/conspiracy} is unevenly distributed across the week, with high activity levels on Sunday and lower activity levels on Monday. The shift in activity suggests that engagement with conspiracy-related discourse fluctuates over time. This fluctuation highlights the importance of temporal context when interpreting engagement, as changes may reflect shifts in attention or availability rather than stable interest.

\textbf{\textit{$\rhd$ Exploratory Insight 3: Cross-Community Content Duplication}} A cross-subreddit comparison between \texttt{r/conspiracy} and \texttt{r/politics} reveals instances of content duplication. The identical content appears in both communities within a short window of time and corresponds to current events, indicating that politically relevant posts circulate rapidly across otherwise distinct subreddits.

\section{Conclusion}
SocialPulse provides an interactive, open-source pipeline for exploratory analysis of Reddit communities, enabling rapid sensemaking through integrated modeling, filtering, and visualization. The system empowers researchers to conduct rapid, exploratory sensemaking of online community discourse by providing an integrated pipeline that unifies data collection, analysis, and interactive visualization.

\section{Acknowledgments}
We thank Majid Alfifi and Haoran Liu for the discussion.
AI tools were used to assist with portions of the coding and to help organize and refine writing ideas. All outputs were reviewed, verified, and modified by the authors.

\bibliography{aaai2026}

\section{Paper Checklist}

\begin{enumerate}

\item For most authors...
\begin{enumerate}
    \item  Would answering this research question advance science without violating social contracts, such as violating privacy norms, perpetuating unfair profiling, exacerbating the socio-economic divide, or implying disrespect to societies or cultures?
    \answerYes{Yes, this work advances understanding of online discourse using publicly available data without violating social contracts.}
  \item Do your main claims in the abstract and introduction accurately reflect the paper's contributions and scope?
    \answerYes{Yes, see Abstract and Introduction.}
   \item Do you clarify how the proposed methodological approach is appropriate for the claims made? 
    \answerYes{Yes, see System Architecture and Implementation.}
   \item Do you clarify what are possible artifacts in the data used, given population-specific distributions?
    \answerYes{Yes, see Introduction and Case Study.}
  \item Did you describe the limitations of your work?
    \answerYes{Yes, see Future Work.}
  \item Did you discuss any potential negative societal impacts of your work?
    \answerNo{No, we did not explicitly discuss potential negative societal impacts because SocialPulse is an only intended to be an exploratory data analysis tool that uses public data.}
      \item Did you discuss any potential misuse of your work?
    \answerNo{No, we did not explicitly discuss potential misuse because SocialPulse is only intended for transparent research and sensemaking.}
    \item Did you describe steps taken to prevent or mitigate potential negative outcomes of the research, such as data and model documentation, data anonymization, responsible release, access control, and the reproducibility of findings?
    \answerYes{Yes, see Conclusion, Bot Filtration, and SocialPulse Dashboard.}
  \item Have you read the ethics review guidelines and ensured that your paper conforms to them?
    \answerYes{Yes}
\end{enumerate}

\item Additionally, if your study involves hypotheses testing...
\begin{enumerate}
  \item Did you clearly state the assumptions underlying all theoretical results?
    \answerNA{NA}
  \item Have you provided justifications for all theoretical results?
    \answerNA{NA}
  \item Did you discuss competing hypotheses or theories that might challenge or complement your theoretical results?
    \answerNA{NA}
  \item Have you considered alternative mechanisms or explanations that might account for the same outcomes observed in your study?
    \answerNA{NA}
  \item Did you address potential biases or limitations in your theoretical framework?
    \answerNA{NA}
  \item Have you related your theoretical results to the existing literature in social science?
    \answerNA{NA}
  \item Did you discuss the implications of your theoretical results for policy, practice, or further research in the social science domain?
    \answerNA{NA}
\end{enumerate}

\item Additionally, if you are including theoretical proofs...
\begin{enumerate}
  \item Did you state the full set of assumptions of all theoretical results?
    \answerNA{NA}
	\item Did you include complete proofs of all theoretical results?
    \answerNA{NA}
\end{enumerate}

\item Additionally, if you ran machine learning experiments...
\begin{enumerate}
  \item Did you include the code, data, and instructions needed to reproduce the main experimental results (either in the supplemental material or as a URL)?
    \answerNA{NA}
  \item Did you specify all the training details (e.g., data splits, hyperparameters, how they were chosen)?
    \answerNA{NA}
     \item Did you report error bars (e.g., with respect to the random seed after running experiments multiple times)?
    \answerNA{NA}
	\item Did you include the total amount of compute and the type of resources used (e.g., type of GPUs, internal cluster, or cloud provider)?
    \answerNA{NA}
     \item Do you justify how the proposed evaluation is sufficient and appropriate to the claims made? 
    \answerNA{NA}
     \item Do you discuss what is ``the cost`` of misclassification and fault (in)tolerance?
    \answerNA{NA}
  
\end{enumerate}

\item Additionally, if you are using existing assets (e.g., code, data, models) or curating/releasing new assets, \textbf{without compromising anonymity}...
\begin{enumerate}
  \item If your work uses existing assets, did you cite the creators?
    \answerYes{Yes, see References and Table 1.}
  \item Did you mention the license of the assets?
    \answerNo{No, the paper does not explicitly list the licenses of external software assets used in SocialPulse. }
  \item Did you include any new assets in the supplemental material or as a URL?
    \answerYes{Yes, see Table 1.}
  \item Did you discuss whether and how consent was obtained from people whose data you're using/curating?
    \answerNA{NA}
  \item Did you discuss whether the data you are using/curating contains personally identifiable information or offensive content?
    \answerNo{No, we do not explicitly discuss the presence of personally identifiable information or offensive content because SocialPulse retrieves publicly available data and analyzes content at an aggregate level.}
\item If you are curating or releasing new datasets, did you discuss how you intend to make your datasets FAIR (see \citet{fair})?
\answerNA{NA}
\item If you are curating or releasing new datasets, did you create a Datasheet for the Dataset (see \citet{gebru2021datasheets})? 
\answerNA{NA}
\end{enumerate}

\item Additionally, if you used crowdsourcing or conducted research with human subjects, \textbf{without compromising anonymity}...
\begin{enumerate}
  \item Did you include the full text of instructions given to participants and screenshots?
    \answerNA{NA}
  \item Did you describe any potential participant risks, with mentions of Institutional Review Board (IRB) approvals?
    \answerNA{NA}
  \item Did you include the estimated hourly wage paid to participants and the total amount spent on participant compensation?
    \answerNA{NA}
   \item Did you discuss how data is stored, shared, and deidentified?
   \answerNA{NA}
\end{enumerate}
\end{enumerate}

\appendix

\begin{table*}[t]
\centering
\small
\caption{\textbf{Comparison With Related Tools}: \Yes indicates yes, \Partial indicates partial, and \No indicates no.}
\label{tab:related-tools}
\begin{tabularx}{\textwidth}{lllllll X}
\toprule
\textbf{Tool} & 
\textbf{Open-Source} & 
\textbf{Reddit} & 
\textbf{Bots} & 
\textbf{Topic Modeling} & 
\textbf{Summarization} &
\textbf{Sentiment} & 
\textbf{Link} \\
\midrule
Hootsuite &
\No & 
\Partial & 
\No & 
\No & 
\Yes & 
\Yes & 
\url{https://www.hootsuite.com/} \\

SproutSocial &
\No & 
\Yes & 
\No & 
\No & 
\Yes & 
\Yes & 
\url{https://sproutsocial.com/} \\

Meltwater &
\No & 
\Yes & 
\Yes & 
\Yes & 
\Yes & 
\Yes & 
\url{https://www.meltwater.com/en} \\

Sanjaya &
\Yes& 
\No & 
\No & 
\No & 
\No & 
\Yes & 
\url{https://github.com/Sanjaya-OSSMM/Sanjaya} \\

\bottomrule
\end{tabularx}
\end{table*}

\section{SocialPulse: System Architecture and Implementation}

\begin{table*}[t]
\centering
\caption{\textbf{Implementation details for SocialPulse processing pipeline shown in Figure~\ref{fig:pipeline},} detailing the tools and stages used for collecting Reddit data, filtering bot activity, extracting topics and sentiment, and presenting results via an interactive dashboard.}
\begin{tabularx}{\textwidth}{@{}p{3cm}p{8cm}X@{}}
\toprule
Tool & Purpose & Link \\

\toprule
\multicolumn{3}{c}{\cellcolor{rowpurple}\textbf{(i) Input \& Data Sources}} \\
\toprule

\textbf{Reddit API} & 
Used to collect publicly available posts, comments, and metadata from user-specified subreddits for analysis. & 
\url{https://www.reddit.com/dev/api/} \\

\toprule
\multicolumn{3}{c}{\cellcolor{rowblue}\textbf{(ii) Bot Filtration}} \\
\toprule

\textbf{BotBuster} \cite{ng2023botbuster} & 
A neural network-based framework that leverages user metadata and posting behavior to estimate bot likelihood in social media accounts.& 
\url{https://github.com/quarbby/BotBuster-Universe} \\

\toprule
\multicolumn{3}{c}{\cellcolor{roworange}\textbf{(iii) Run Analysis}} \\
\toprule

\textbf{BERTopic} \cite{grootendorst2022bertopic} & 
A transformer-based framework that utilizes Sentence-BERT embeddings and clustering (HDBSCAN) to discover latent themes in unstructured Reddit text. & 
\url{https://maartengr.github.io/BERTopic/index.html} \\

\textbf{LLM Tooling} & 
LLMs are integrated to refine topic representations, generate concise labels, and provide higher-level summaries of complex threads. We use \texttt{gpt-4.1-mini}. & 
\url{https://platform.openai.com/docs/api-reference/introduction} \\

\textbf{VADER Sentiment Analysis} \cite{hutto2014vader} & 
A lexicon-based sentiment analysis tool that computes the polarity of posts and comments.& 
\url{https://pypi.org/project/vaderSentiment/} \\

\toprule
\multicolumn{3}{c}{\cellcolor{rowyellow}\textbf{ (iv) SocialPulse Dashboard}} \\
\toprule

\textbf{Flask} & 
A lightweight web framework used to manage the backend routing and data exchange of the interactive SocialPulse dashboard. & 
\url{https://flask.palletsprojects.com/en/stable/} \\

\bottomrule
\end{tabularx}
\label{tab:links}
\end{table*}

SocialPulse is designed as a modular, end-to-end pipeline. The system architecture is divided into four sequential modules: Data Ingestion, Bot Filtration, Analytical Engine, and the Interactive Visualization Interface. We provide an overview of the pipeline in Figure~\ref{fig:pipeline}, and the detailed links in Table~\ref{tab:links}.

\subsection{(i) Data Collection and Statistics}
SocialPulse leverages the Reddit API to fetch public data based on user input. Users may analyze one or multiple subreddits and configure data collection by either specifying a fixed number of posts retrieved via the Reddit API using the \textit{Best, Hot} and \textit{Recent} sorting categories or by constraining the data to a particular time interval. This enables a flexible exploration of both ``high-traffic'' and ``long-tail'' community dynamics. Consider \texttt{r/AskHistorians}, for example. Although posts to this subreddit often receive only a few responses, the responses are more-depth and higher-effort compared with other subreddits. As is typical in niche communities, interactions between community members may be less frequent, and so a longer time interval is necessary to observe the network structure. Tailoring the data collection method towards the community of interest based on group sizes and time intervals is important \cite{panek2018effects}.

\subsection{(ii) Bot Filtration}
To ensure the integrity of social media analysis, the system includes an optional bot detection stage. We utilize the BotBuster tool \cite{ng2023botbuster}, a mixture-of-experts neural network architecture designed for multi-platform bot detection. BotBuster analyzes information pillars such as user metadata and posts to estimate the probability of an account being a bot. This allows researchers to isolate human-driven discourse from automated amplification.

\subsection{(iii) Analytical Engine}
Once the data is cleaned, SocialPulse runs a suite of complementary analyses to extract thematic and behavioral patterns:

\begin{itemize}

\item BERTopic Analysis \cite{grootendorst2022bertopic}: A transformer-based framework that utilizes Sentence-BERT embeddings and clustering (HDBSCAN) to discover latent themes in unstructured Reddit text.

\item LLM Tooling: Large Language Models are integrated to refine topic representations, generate concise labels, and provide higher-level summaries of complex threads.

\item VADER Sentiment Analysis \cite{hutto2014vader}: A lexicon-based sentiment analysis that computes the polarity of posts and comments.

\item Cross-Community Comparison: The system supports comparative analysis across multiple user-specified subreddits that examines shared themes, duplicate posts, and overlapping users. 

\end{itemize}

\subsection{(iv) SocialPulse Dashboard}

The final output is a user-centralized interactive dashboard. It provides a suite of features for rapid sensemaking:

\begin{itemize}
\item Exploratory Topic Analysis: Interactive visualizations of the BERTopic clusters present topic labels, representative keywords, and example posts, allowing users to drill down from aggregate trends to individual posts and raw comments for contextual analysis.

\item Discourse Comparison: Side-by-side comparisons of sentiment polarity distributions, user activity levels, and topic frequencies across subreddits support analysis of differences in discourse tone and thematic emphasis among communities.

\item Bot Transparency: Visual indicators of bot probability and flagged content, enabling users to evaluate the extent to which automated users contribute to observed discourse patterns.

\item Configurable Analysis: The interface allows users to customize the analysis by adjusting BERTopic hyperparameters, specifying one or multiple subreddits, setting the minimum threshold required to flag an account as a bot, and selecting between a fixed-size or time-based data collection.
\end{itemize}

SocialPulse bridges the gap between raw data collection and deep qualitative understanding. By open-sourcing the toolkit, we aim to provide the ICWSM community with a reproducible foundation for studying social media dynamics.

\section{Broader Impact}

Computational social science research increasingly relies on large-scale social media data, yet the analytical workflows required to interpret this data remain fragmented across disparate tools and scripts. SocialPulse aims to support researchers by providing an integrated, interactive environment that connects data collection, modeling, and exploratory analysis within a single workflow to improve exploration and hypothesis generation.

A key contribution of SocialPulse is its support for examining participation structure alongside topical and sentiment-based analyses. By explicitly distinguishing between highly active users and long-tail contributors, the toolkit helps researchers avoid over-reliance on highly active users and encourages more nuanced interpretations of online discourse. This is particularly valuable for studies concerned with community norms and the dynamics of attention and engagement.

More broadly, SocialPulse is intended to improve transparency and reproducibility in exploratory social media analysis. By offering an open-source pipeline with clearly defined analytical components, the system allows researchers to inspect and adjust parameters, and replicate exploratory workflows across datasets and communities. In doing so, SocialPulse supports more replicable sensemaking practices in computational social science, rather than replacing or automating interpretation.

\section{Future Work}

Currently, SocialPulse is optimized for Reddit's nested comment structure. Future iterations will include:

\begin{itemize}
    \item \textbf{Cross-Platform Support:} Extending the pipeline to support other social media platforms.
    \item \textbf{LLM-in-the-Loop:} Integrating LLMs to provide automated ``Sensemaking Summaries'' of complex thread hierarchies.
    \item \textbf{Real-time Notifications:} Enabling practitioners to set thresholds for ``sentiment spikes'' or ``bot activity'' in monitored communities.
\end{itemize}

\end{document}